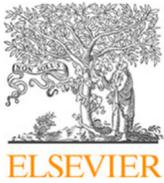
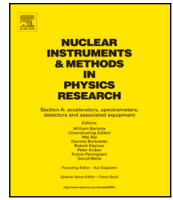

# Development of a real-time beam profile monitor for GeV photons and its application in accelerator facilities

R. Kino [a,b],*, S. Nagao [c], T. Akiyama [a,b], H. Fujioka [d], T. Fujiwara [a], T. Ishige [a,b], K. Itabashi [e], S. Kajikawa [a,b], M. Kaneta [a], M. Mizuno [a], S.N. Nakamura [c,a], K. Nishi [c], K. Nishida [c], K. Okuyama [a,b], F. Oura [a,b], K. Tachibana [a,b], Y. Toyama [f], D. Watanabe [a]

[a] *Graduate School of Science, Tohoku University, Miyagi, 980-8578, Japan*
[b] *Graduate Program on Physics for the Universe (GP-PU), Tohoku University, Miyagi, 980-8578, Japan*
[c] *Graduate School of Science, The University of Tokyo, Tokyo, 113-0033, Japan*
[d] *School of Science, Institute of Science Tokyo, Tokyo, 152-8550, Japan*
[e] *International Center for Quantum-field Measurement Systems for Studies of the Universe and Particles, Ibaraki, 305-0801, Japan*
[f] *Center for Muon Science and Technology, Chubu University, Aichi, 487-8501, Japan*

## ABSTRACT

A real-time beam profile monitoring system is proposed for GeV photon beams at the BM4 beamline of the Mikamine site, Research Center for Accelerator and Radioisotope Science (RARiS; previously known as ELPH) at Tohoku University. This monitoring system enhances the capability to monitor the entire beamline by incorporating newly developed beam profile monitors (BPMs) for upstream and midstream sections, in addition to the existing high-speed BPM used for downstream monitoring. This paper reports on the detection mechanisms of the newly developed BPMs and the actual measurement results obtained using the integrated beam monitoring system. The new BPMs are composed of plastic scintillation fibers and silicon photomultipliers, enabling high-precision, real-time measurements. Data acquisition utilizes streaming TDC, a firmware commonly employed in the J-PARC Hadron-hall, allowing real-time detection of high-intensity photon beams with count rates reaching several tens of MHz. With sufficient statistical data, the BPM achieved a 1-s beam-profiling accuracy of 10 μm. The proposed BPM system serves as a valuable resource for future physics experiments at the BM4 photon beamline and will significantly contribute to ongoing accelerator research endeavors.

## 1. Introduction

In several physics experiments using particle accelerators, gaining a comprehensive understanding of the properties of the beams generated by the accelerators is crucial. Such understanding facilitates the collection of higher-quality physics data and improves the tuning and performance of the accelerators.

At the Mikamine site, Research Center for Accelerator and Radioisotope Science (RARiS; previously known as ELPH) at Tohoku University, the electron synchrotron BST ring produces bremsstrahlung photon beams in the 1 GeV region [1]. The BST ring accelerates electrons initially injected with an energy of 90 MeV up to 1.3 GeV, and they are then stored within the ring. The injection cycle lasts approximately 17 s; during typical operation, a flat-top phase of approximately 10 s is established [2].

To generate a bremsstrahlung photon beam, a $\phi$11 μm carbon fiber radiator is inserted into the path of orbiting electrons during the flat-top phase. This radiator synchronously moves along with the accelerator's operation and remains stationary at a location distant from the electron orbit. As the electrons are injected and accelerated within the BST ring, the radiator gradually shifts toward the center of the electron intensity distribution on the flat-top phase. Following the conclusion of the flat-top phase, the radiator is immediately removed from the electron orbit, with its precise operation controlled by a stepping motor [3].

The BST ring accommodates two photon beamlines, namely, BM4 and BM5. BM4 beamline is equipped with an electromagnetic spectrometer, NKS2 [4], while BM5 beamline has an electromagnetic calorimeter known as FOREST [5]. Both beamlines are equipped with photon energy tagging devices referred to as Tagger [3,6]. The Tagger can identify the energy of the photon beam by measuring the timing and position of the scattering electrons generated by bremsstrahlung, all without disrupting the photons.

The 1 GeV region photon beam, tagged using this method, proves useful for investigating hadrons, including strangeness nuclear physics, and several experiments have been conducted on this beamline. Accurate beam profiling is necessary for several reasons, such as measuring reaction points and designing the positions of detectors and targets. Additionally, the accuracy of the beam profile significantly affects the final physical results.

Previous methods employed for profiling photon beams at the BM4 beamline have proven insufficient in meeting the accuracy requirements of these physical experiments. Two distinct tools have been used to ascertain the beam profile:

The first method involves the use of instant camera film (Fujifilm, Instax), operating on the principles of an emulsion plate. In this method, the beam is directed onto the film, and a photograph is developed using a camera subsequent to stopping the extraction beam and the operation of the BST ring. However, this method suffers from the following disadvantages: (1) it lacks the capability to quantitatively measure the beam profile, and (2) it requires a considerable amount of time to visualize the profile.

The second method uses the high-speed beam position monitor (HSBPM [7]), which is a detector employing scintillation fibers (Sci-Fi) (Saint-Gobain, BCF-10SC, $\phi 3.0$ mm) and a multi-anode photomultiplier tube (PMT; Hamamatsu Photonics K.K., H6568-10MOD). This HSBPM incorporates two types of plastic scintillators for the VETO counter of charged particle backgrounds and the trigger counter, an aluminum photon converter for generating electron–positron pairs from a portion of the photon beam, and Sci-Fi layers for precise particle position determination. While this detector offers the ability to quantitatively measure the photon-beam profile, its size, and Sci-Fi segments limit its usage exclusively to the downstream (at a distance of approximately 10 m) of the beamline.

Therefore, in the present study, a new photon-beam profile monitoring system was developed for this beamline. A new beam profile monitor (BPM) with detector and segment sizes suitable for upstream and midstream beam monitoring was devised, and it can be used in combination with the HSBPM at three different points in the beamline: upstream, midstream, and downstream. Furthermore, it is necessary for the real-time monitoring and quantitative evaluation. For this reason, a system with a fast response time was introduced. For instance, the use of plastic scintillators with silicon photomultiplier (SiPM) readout has been successfully demonstrated in various applications, such as cosmic ray imaging systems [8] and light trackers for particle detection [9]. Inspired by these advancements, the newly developed BPM employs scintillating fibers and SiPMs. A streaming readout system was also incorporated to efficiently measure and real-time monitor high-intensity beams. In this study, we performed an assessment of the proposed BPM, considering that it was specifically designed for high-energy and high-intensity photon beams, based on measurement results obtained using the entire monitoring system.

## 2. System for detecting photons that are neutral particles

The primary structure of the BPM comprises Sci-Fi and SiPMs, employed for the detection of photon beams composed of neutral particles. The detection component of the BPM features a five-layer structure, as illustrated in the schematic in Fig. 1. This section provides an explanation for each layer.

### 2.1. Charged VETO counter for removing charged backgrounds

The upstream counter within the beam serves as a VETO counter designed to detect charged particles, which represent background events within the laboratory. These charged particles, such as electron–positron pairs formed when the photon beam interacts with materials including air and the aluminum flange at the radiator position, can distort the beam profile if not appropriately considered. Therefore, effectively mitigating this background with the charged VETO counter is essential to accurately profile the photon beam. However, excessive material thickness must be avoided as it can lead to unnecessary electron–positron pair production within the detector, which would suppress the number of detected photon events.

To attain a balance between detection efficiency and material thickness, we used 99 scintillation fibers (Kuraray, SCSF-78) with a diameter of 0.5 mm for the charged VETO counter, arranged in double layers to eliminate gaps. This design ensures that the VETO counter can efficiently detect charged particle background while maintaining a minimal effective thickness of approximately 0.93 mm in the beam direction. A photon conversion probability is approximately 0.17% for this thickness. The Sci-Fi were bundled in groups of 33 and connected to three multipixel photon counters (MPPCs) (Hamamatsu K.K., S13360-3050PE) linked in parallel.

The impact of the charged VETO efficiency on the beam profile was estimated by comparing the profiles of events detected and not detected by the charged VETO counter during offline analysis. We found that the charged VETO counter's detection efficiency does not have a significant impact on the beam position. The two values agree with an error below 10 $\mu$m. On the other hand, as for the beam size, the events detected by the charged VETO counter have approximately 11% larger size than those not detected, in case of position at 2.99 m downstream from the radiator. Therefore, depending on the VETO counter detection efficiency, the beam size may be slightly overestimated, particularly owing to the spread of charged particle background events (e.g., electrons and positrons generated upstream). To estimate the impact of the VETO counter detection efficiency on the measured photon beam size, assuming an extreme case where the detection efficiency is 50%, the photon beam size would be approximately 2% smaller than the measured results, and the beam halo area would be approximately 9% smaller.

### 2.2. Aluminum photon converter

The subsequent layer is the photon converter, consisting of a thin aluminum plate with a thickness of 430 $\mu$m. Its primary function is to generate electron–positron pairs from a small fraction (0.38%) of the photon beam. The ensuing Sci-Fi layers immediately detect the generated electron–positron pairs to determine the position of the incident particle. Minimizing the distance between the photon converter and Sci-Fi layers is crucial for mitigating the effect of multiple hits. To this end, the distance between the photon converter and Sci-Fi layers was reduced to 0.5 mm or less.

### 2.3. Fiber layers for determining particle position

The next two layers consist of Sci-Fi fibers with a diameter of 0.5 mm (Kuraray, SCSF-78). These layers comprise 45 fibers arranged in both the direction of gravity ($x$ channel) and horizontal direction ($y$ channel), serving the purpose of determining the positions of particles within the beam. The use of 0.5 mm diameter fibers helps in reducing the effective thickness of the detector along the beam axis, thereby suppressing excessive electron-positron pair production.

To guarantee uniformity in signal response across the detector plane, we evaluated the uniformity by examining how the measurement results shifted while moving the position of the BPM itself using a stepping motor (with a repeated crossover of 3 $\mu$m). The results confirmed that the beam center position coincides with the motor's moving distance with an error of less than 10 $\mu$m, indicating consistent signal response across channels. The variations in light transmission, production, and coupling efficiency to the SiPMs between channels were evaluated and confirmed to have no significant impact on the measurement results.

At the end of each fiber, three fibers are bundled together, and their tips make contact with the MPPCs (Hamamatsu K.K., S13360-1350PE).

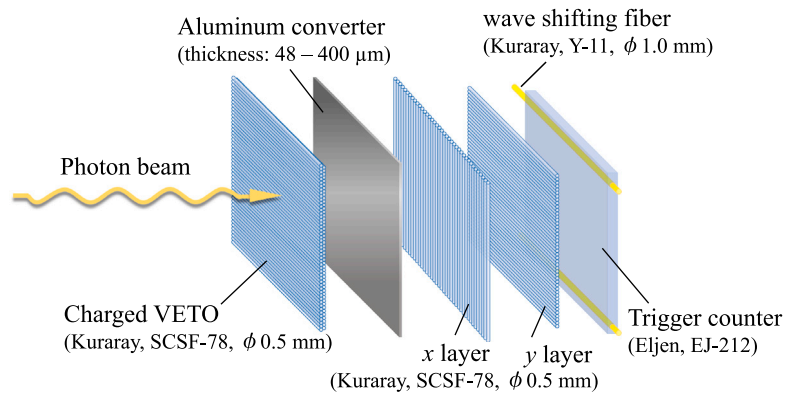

**Fig. 1.** Schematic of the particle detection portion composed of five layers in order from the upstream of the beam. The first layer is the VETO counter (Section 2.1) used to detect the background of charged particles. The second layer is the aluminum photon converter (Section 2.2) responsible for generating electron–positron pair production from a part of the photon beam. The third and fourth layers are the Sci-Fi layers (Section 2.3) used to determine the particle position. The fifth layer is the trigger counter (Section 2.4) used for event identification. The effective area for detection is 22.5 × 22.5 mm$^2$.

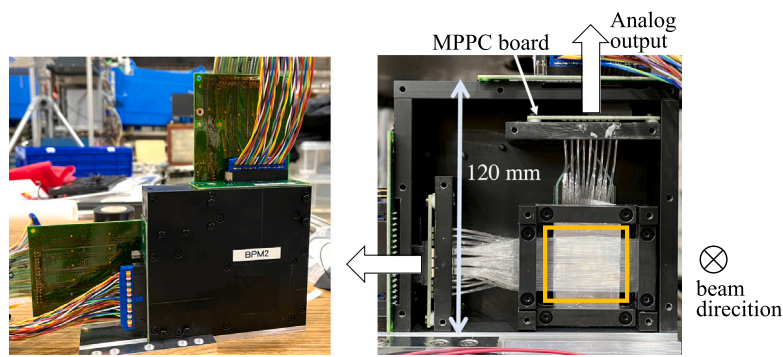

**Fig. 2.** Exterior (left) and interior (right) of the developed BPM. The frame is made of black ABS resin, which is easy to process and has excellent impact resistance. The detection part is the part surrounded by the square in the right figure. Each MPPCs are fixed in contact with the end face of each fiber. The MPPC's analog output signals are sent to the signal processing circuit (Section 3) outside the frame.

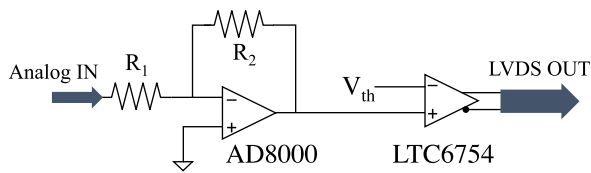

**Fig. 3.** Schematic diagram of a signal amplifier circuit for one channel. The small analog signal output from the MPPC is amplified by a non-inverting amplifier circuit that includes the AD8000 operational amplifier. It is then converted into a digital differential signal via the high-speed comparator, LTC6754.

Each channel has a width of 1.5 mm, with a total of 15 channels in both vertical and horizontal directions. This configuration ensures sufficient efficiency and accuracy for profiling a photon beam with a sigma value ($\sigma$) of approximately 1.5 mm.

### 2.4. Trigger counter for event identification

The final layer within the detection section is the trigger counter, responsible for event identification. It employs a 2.0 mm thick plastic scintillator plate (Eljen, EJ-212) to ensure that no events are missed. Wavelength shifting fibers (Kuraray, Y-11, $\phi$1.0 mm) are embedded at both ends of the scintillator plate, and the resulting scintillation light is detected by two MPPCs (Hamamatsu K.K., S13360-1350PE) connected in parallel.

All counters are connected to the MPPCs, detecting the scintillation light as a signal, which is then processed into a digital signal (as described in Section 3.1) and collected using the Hadron Universal Logic (HUL) time-to-digital converter (TDC), referred to as streaming TDC (see Section 3.2). Once all the data are recorded, events are identified offline by applying the conditions outlined in the following formula:

$$\text{Photon event} = \overline{[\text{Charged VETO}]} \otimes [x \text{ layer}] \otimes [y \text{ layer}] \otimes [\text{Trigger}] \quad (1)$$

An event is defined as the time difference between the trigger counter and the $x$- and $y$-channels, with the charged VETO counter falling within the specified time window of 48 ns. This method ensures the effective detection of neutral-photon beams.

The complete five-layer detector system was securely mounted on a specially designed acrylonitrile butadiene styrene (ABS) resin frame, as depicted in Fig. 2. An original circuit board was developed to align the MPPCs, which read the signals, with the fiber spacing. Both the board and fibers were firmly affixed to the frame to ensure stable and accurate measurements.

## 3. Readout and signal processing system

This section describes the readout and signal processing systems for the data obtained from the high-intensity photon beam. A circuit was developed (Section 3.1) to convert the analog output signals from the MPPCs into a format suitable for the data acquisition (DAQ) module (Section 3.2). Section 3.3 provides an overarching view of the DAQ process.

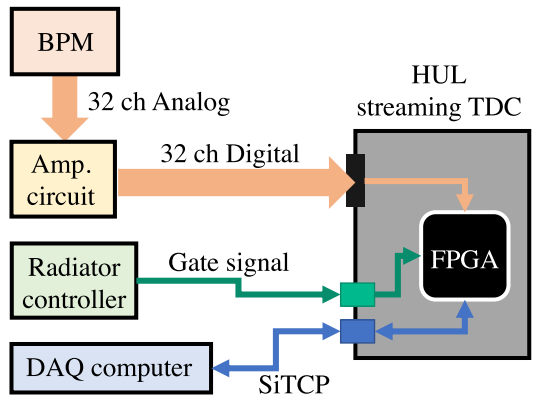

**Fig. 4.** Schematic of the data acquisition system. First, the signal obtained from the BPM is amplified by an amplifier circuit and converted into a digital differential signal, which is then input to the HUL as an LVDS signal. In contrast, the gate signal synchronized with the movement of the radiator is input independently to the HUL from the radiator controller, which enables data acquisition synchronized with the photon beam. The DAQ computer and HUL communicate with each other via SiTCP.

### 3.1. Signal amplifier and converter circuit

To accurately read the signal from the MPPCs, the analog signal must be amplified to an optimal level and convert it into a digital differential signal. For this purpose, a dedicated reading circuit was devised, as shown in Fig. 3. The circuit encompasses an amplifier, employing an operational amplifier (Analog Devices, AD8000YRDZ) to amplify the MPPC signal approximately 20-fold, and a conversion circuit, which employs a high-speed comparator (Linear Technology, LTC6754) to convert the signal into an LVDS signal. A similar circuit was replicated for all 32 channels per BPM.

The digital conversion method utilized the time over threshold (ToT) method, which converts wave height information into time information and measures it. In this method, the digital output contains time information at the leading edge and energy information corresponding to the pulse width. This allows for the simultaneous acquisition of time and energy information using a single signal line, reducing the circuit size of the DAQ system and simplifying the configuration, ultimately reducing the overall cost. The HUL firmware (Section 3.2), used as the DAQ module, accepts event durations within 4–150 ns. To maximize the use of this time range, we shaped the analog signal such that the median pulse width was approximately 80 ns.

The digital conversion comparator LTC6754 incorporates internal hysteresis, which can be adjusted within the range of 0–40 mV based on the value of the connected external resistor. In our developed conversion circuit, the hysteresis was configured at 40 mV. The output corresponding to one photoelectron (one pixel of the MPPC) was about 45 mV. For the $x$ and $y$ layers, the digital conversion threshold voltage ($V_{th}$) was set to 280 mV, corresponding to 5–6 p.e. (photoelectrons), approximately half the average photon number expected for this detector system. The average photon count for this detector was measured and estimated in a beam test using $\phi = 1.5$ mm fiber (Kuraray SCSF-78) and the same type of MPPC (Hamamatsu Photonics K.K., S13360-1350PE) as the detector. Given the high background level in the beam-irradiated environment, we determined the optimal threshold by scanning the $V_{th}$ and observing the beam profile during an actual in-beam test. While it is true that the threshold of 5–6 p.e. could potentially reduce detection efficiency, our tests confirmed that this did not significantly impact the uniformity of the measurements or the overall experimental results, as mentioned in Section 2.3. $V_{th}$ for the Charged VETO and Trigger counters were set to 70 and 90 mV, respectively, to prevent oscillations due to noise.

### 3.2. Streaming TDC

The primary module used for DAQ was the HUL [10], an FPGA module commonly used in recent nuclear experiments. We implemented a streaming TDC [11], which facilitated triggerless continuous timing measurements. This firmware achieved a high dynamic range of approximately $10^{10}$ timing measurements using the heartbeat method, which periodically generated delimiter data. Each TDC channel had a width of 0.96 ns, and it could collect continuous data for up to 32 s, rendering it suitable for profiling high-intensity photon beams with a counting rate of several MHz.

The integration of the streaming TDC enabled us to measure the tagged photon beams at counting rates of several MHz. Since all counter timestamps can be recorded without selection by online triggers, the desired data can be chosen after the DAQ is complete, as shown in Eq. (1).

### 3.3. Data acquisition system

Fig. 4 presents a block diagram that outlines the entire process of DAQ. Importantly, the gate signal needed to be supplied separately to the HUL, utilized for DAQ, from the data signal. The streaming TDC continued to acquire data only when the gate signal was active. To synchronize with the photon beam, we employed a gate signal that became active when the internal radiator was in motion. Communication between the HULs and computers for DAQ occurred through Ethernet using SiTCPs technology [12]. In the actual test experiment, one computer simultaneously controlled four HULs, and DAQ was performed accordingly.

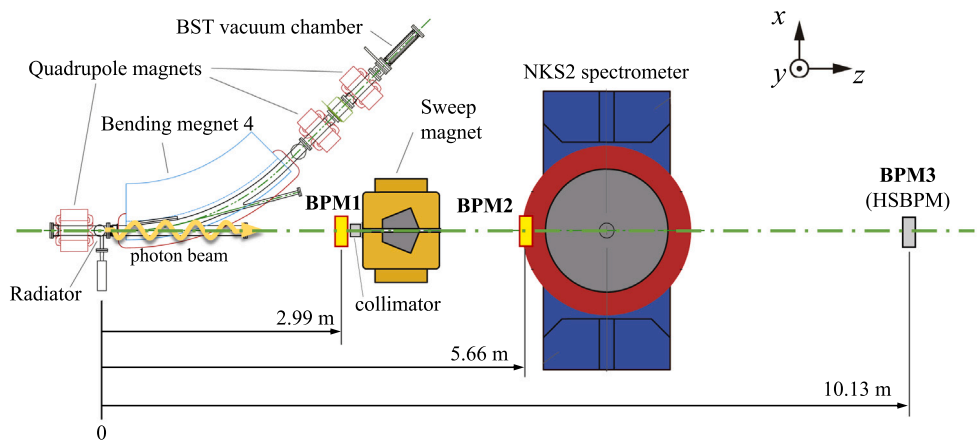

**Fig. 5.** Schematic view of the BM4 beamline. The photon beam is generated at the radiator, located at the entrance of the dipole magnet, known as Bending Magnet 4. To minimize electron-positron background events, a Sweep magnet is also installed in the beamline. We have installed our newly developed Beam Profile Monitors (BPM1 and BPM2) in the up- and the midstream of the beamline. The BPM3(HSBPM) is placed at the most downstream point of the beamline.

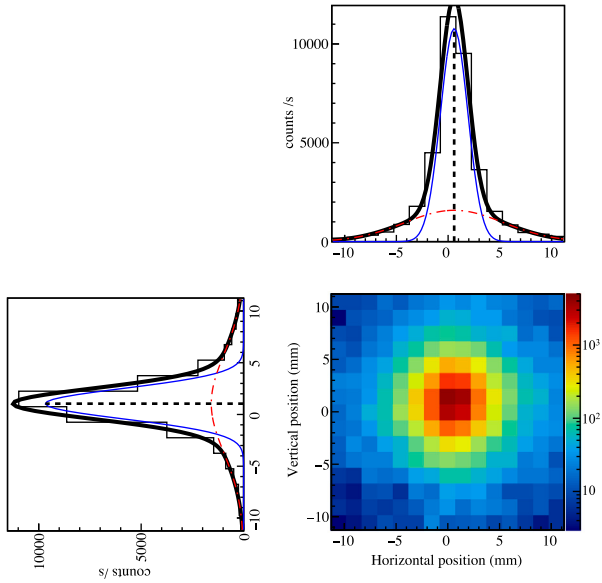

**Fig. 6.** Two-dimensional distribution of hits in the *x*–*y* plane over a 1-s period. The hits in the *x*-layer, corresponding to the horizontal direction, and the *y*-layer, corresponding to the vertical direction, are integrated and plotted. The blue line represents the real beam structure, while the red line represents the beam halo structure. The fitted function is a double Gaussian, represented by the black line, indicating the sum of the values corresponding to the blue and red lines.

## 4. Measured photon profiles

### 4.1. Setup for the test experiment

The experimental setup used in this study is illustrated in Fig. 5. The coordinate system within the BM4 photon beamline was established with the *y*-axis pointing opposite to the gravitational direction, the *z*-axis aligned with the direction of the photon beam, and the *x*-axis perpendicular to both the *y* and *z*-axes. The *x*-axis pointed toward the inside of the BST ring, following the right-handed system convention. Two new BPMs, BPM1 and BPM2, were developed and installed on the beamline to monitor the photon beams. BPM1 possesses the capability to move in the *x* and *z*-directions using two movable stages and can be remotely controlled to position in front or behind the collimator. BPM2 remained fixed at the entrance of the NKS2 spectrometer.

Additionally, the existing HSBPM was installed as BPM3 downstream of the beamline. The entire DAQ system for BPM3 was replaced by a new system, incorporating a streaming TDC, akin to BPM1 and 2. The analog output signal from the multi-anode PMT was initially converted to an emitter coupled logic (ECL) signal utilizing an NIM standard discriminator (GeV$\gamma$ – 1380). The digital conversion threshold voltage, $V_{th}$, was set to 40 mV, which corresponds to approximately one-fifth of the signal gain associated with the minimum energy loss (minimum ionization particle (MIP)) of a general plastic scintillator. Subsequently, the HUL collected data in the same manner as BPM1 and 2.

The entire DAQ system comprised three HULs, each equipped with a StrTDC, independently reading 32 channels for each BPM. A gate signal synchronized with the movement of the radiator was simultaneously input to all HULs. A DAQ computer controlled all the HULs via SiTCP. The same DAQ computer also handled radiator control and BPM1 stage control. The typical data size for each HUL was as follows: BPM1: approximately 75 MB/spill, BPM2: approximately 55 MB/spill, BPM3: approximately 400 MB/spill. The typical DAQ rates for each detector were BPM1: approximately 85 kHz, BPM2: approximately 54 kHz, BPM3: approximately 330 kHz (orbiting electron-beam current approximately 1 mA). StrTDC is capable of collecting time stamps at high rates, meaning that only 1 s of DAQ is sufficient to obtain ample statistical data. Therefore, the beam-profiling system developed using these three BPMs can operate in real time immediately after collecting data for one spill.

The distances of these BPM *x*-layers from the radiator were accurately measured using a laser rangefinder and were found to be 2.99, 5.66, and 10.13 m, respectively.

### 4.2. Quantitative evaluation of beam profile

Photon event selection was performed using Eq. (1) for the timestamps of all events recorded by the streaming TDC. The time window was set to ±48 ns, effectively covering the difference in the timestamp between the trigger counter and each layer.

Fig. 6 depicts a typical beam profile, representing a 1-s profiling. The number of hits is presented as a histogram in both the horizontal and vertical directions, and a fitting process was performed using a function that incorporated two Gaussians (depicted using black lines in Fig. 6). The blue line corresponds to the real beam structure, while the red line represents the beam's halo structure. It is conceivable that the origin of this halo structure lies in the charged particles generated at the aluminum flange, influenced by the efficiency of the charged

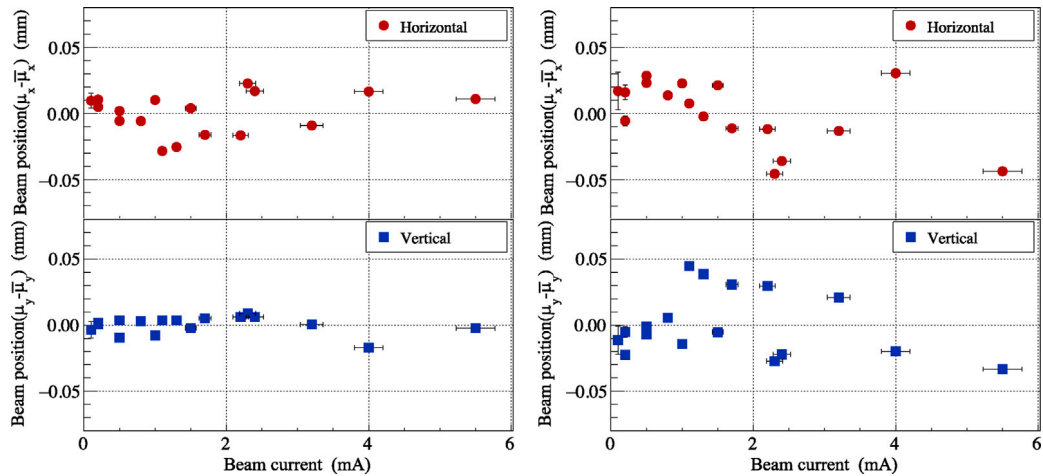

**Fig. 7.** Dependence of the beam position on the beam intensity at two different points along the beamline. The figure on the left represents the position measurements from BPM1, which is located 2.99 m from the radiator, and the figure on the right represents measurements from BPM2, located 5.66 m from the radiator. The *y*-axis was set so that the average value ($\bar{\mu}$) would be 0 for each BPM. The red points in the figures represent the horizontal position of the beam, and the blue points represent the vertical position.

VETO counter. The mean value ($\mu$, represented by the black dotted line) derived from the fitting was considered as the beam center position, while the sigma ($\sigma$) was assessed as the beam size. The typical position accuracy was less than 10 μm for the 1-s profile.

The bias voltage of the SiPMs for $x$ layer and $y$ layer was set uniformly, ensuring consistent operation across all channels. Under this constant bias voltage, the signal gain equivalent to one photoelectron for each channel (the output signal of the operational amplifier AD8000YRDZ after passing through the signal amplification circuit) was evaluated. The gain of the MPPC output signal for each channel remained constant with an accuracy of less than 10%. This gain stability was confirmed by examining the relationship between the number of photons and the signal gain of the MPPC output.

To examine the impact of this gain variation on detection efficiency, we assumed a Poisson distribution of the number of photoelectrons per event. With a mean of Poisson of 11 p.e. and a digital conversion threshold of 6 p.e., only 1% of all detected events can be missed because of MPPC gain variations of 10% when the event actually had a signal above the threshold.

With this 1% effect on detection efficiency, the resulting fluctuation in the final profiling results was only a few micrometers. This fluctuation is deemed negligible and has no significant impact on the values obtained in this study. Additionally, the error estimates for these values encompass solely the statistical errors obtained from chi-square fitting.

### 4.3. Beam intensity dependence

The investigation of the photon-beam profile involved varying the electron-beam current within the BST ring. Since the beam current could not be directly measured, the beam intensity was evaluated using the measurement results obtained from the beam DCCT. Fig. 7 shows the relationship between the beam position and intensity, revealing that there was no significant alteration in the beam position as the beam intensity increased. Fig. 8 portrays the dependency of the beam size on the beam intensity; the beam size exhibited a minor increase as the beam intensity rose. This increase encompassed the effect of accidental background generated by the beam intensity. In scenarios with high beam intensity (count rate = 150 kHz), the beam size is estimated to be approximately 2% larger owing to the influence of this accidental background.

### 4.4. Time dependence during a spill

Fig. 9 shows the time variation of the beam profile during a spill. The plot demonstrates that the direction of the beam undergoes changes within a spill, with the horizontal variation consistent with the characteristics determined by the unique Twiss parameter of the BST ring and the $x-x'$ phase space of the electron beam. The $x-x'$ phase space defines the particle distribution and is typically determined by the Twiss parameters of the accelerator. At the radiator position in the BM4 beamline, the particle intensity within the $x-x'$ phase space exhibits an opposite correlation, leading to the horizontal beam position moving in the direction opposite to that of the radiator's motion.

Furthermore, the plot shows changes in the vertical direction, indicating that the $x-y$ plane of the electron-beam position distribution at the radiator point is elliptically tilted. This phenomenon may be attributed to the alignment of the quadrupole magnets within the BST ring.

Fig. 10 illustrates the time-dependent behavior of the beam size, indicating a significant increase during a spill. This phenomenon is attributed to the degradation of the electron-beam emittance caused by the insertion of a radiator. When the radiator moves within the electron beam, it induces electron scattering, which in turn degrades the emittance and enlarges the size of the photon beam [13]. The lower graph in Fig. 10 shows the time-dependent behavior of the beam size measured using BPM3. This effect of electron scattering was not observed at greater distances from the radiator owing to the larger original beam size.

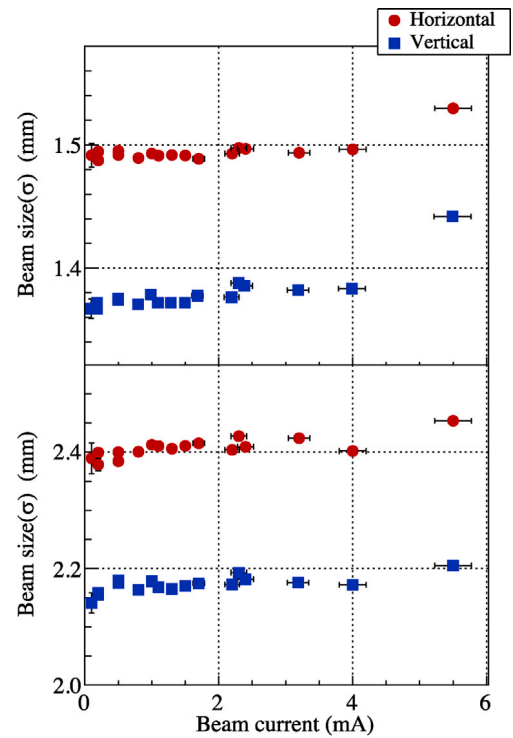

**Fig. 8.** Beam intensity dependence of the beam size, with BPM1 (at a distance of 2.99 m from the radiator) on the upper and BPM2 (at a distance of 5.66 m from the radiator) on the bottom. The red points represent the horizontal beam size, while the blue points represent the vertical beam size.

### 4.5. Radiator position dependence

During the operation of the BST ring, the radiator continually moves within a spill. To assess the dependence of the beam on the radiator position, the radiator was inserted at a specified point, $x_{\rm rad}$, and then fixed in place after the electron beam was accelerated. The initial position of the radiator was recorded on both the outer and inner circumferences of the ring for both patterns. The results for both patterns are presented in Figs. 11, 12, and 13. In these figures, the 0 point on the horizontal axis corresponds to the center of the electron beam.

Fig. 11 depicts the relationship between the beam center position and radiator position. The figure distinctly shows the correlation between the radiator's position and direction of the photon beam. Regarding the horizontal direction, the characteristics explained in the previous section related to the accelerator-specific Twiss parameter and $x-x'$ phase space are clearly evident. A correlation in the vertical direction is also observed, confirming that the photon beam exhibits an upward behavior near the center of the electron beam.

Figs. 12 and 13 show the dependence of the photon-beam size on the radiator position. The photon-beam size notably increases as the radiator approaches the center of the electron beam. The rate of increase in the vertical direction is more substantial because the emittance in the vertical direction is much smaller than that in the horizontal direction. Consequently, the effect of inserting the radiator is more pronounced in the vertical direction owing to the smaller initial size of the beam in that direction [14]. The error bars in the positive radiator position area are larger because the radiator is inserted from the outer side of the BST ring, leading to most of the electron beam being intercepted before reaching the specified point $x_{\rm rad}$, resulting in statistical errors.

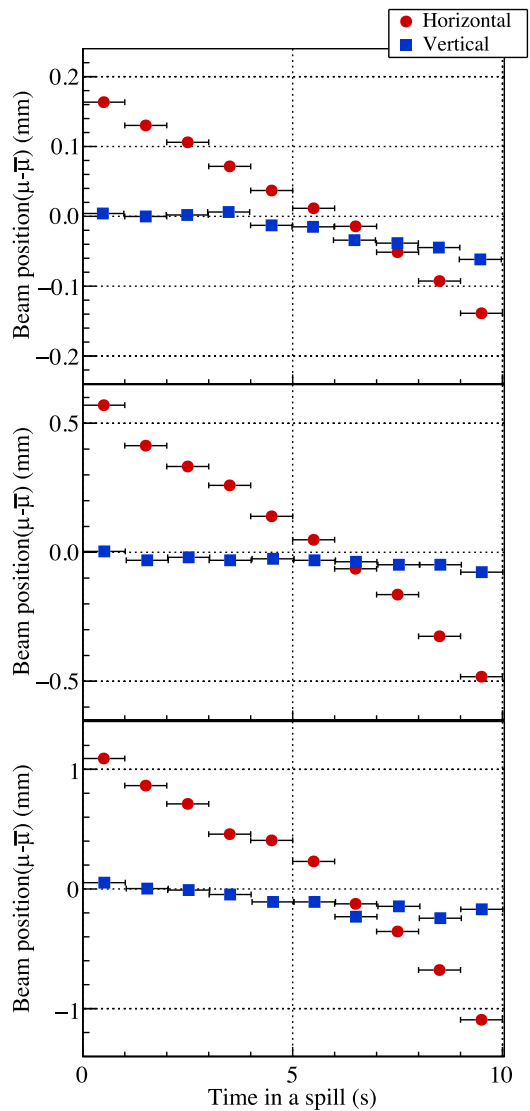

**Fig. 9.** Time dependence of the beam position. The upper figure is measured by BPM1 (the distance from the radiator is 2.99 m), the middle is by BPM2 (the distance from the radiator is 5.66 m), and the bottom one is by BPM3 (the distance from the radiator is 10.13 m), respectively. The red points represent horizontal, and the blue ones represent vertical positions.

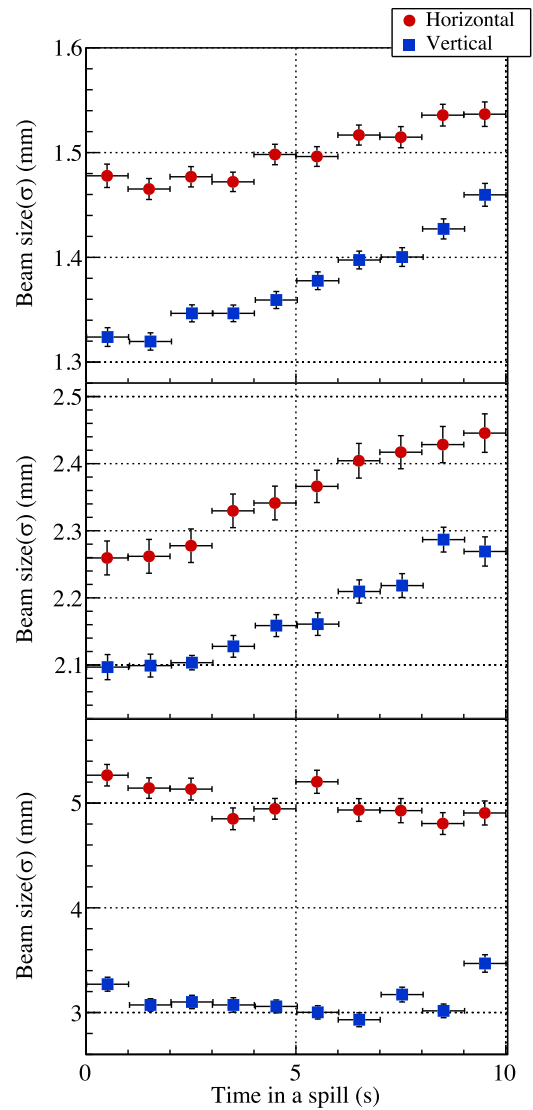

**Fig. 10.** Time dependence of the beam size. The upper figure corresponds to BPM1 (the distance from the radiator is 2.99 m), the middle figure is BPM2 (the distance from the radiator is 5.66 m), and the bottom is measured by BPM3 (the distance from the radiator is 10.13 m), respectively. The red points represent the horizontal beam size, and the blue ones represent the vertical beam size.

During standard operation, the radiator moves within the range of −2.8 to −2.1mm from the outer circumference to the inner circumference of the ring, aligning with the findings obtained from the time-dependent analysis (Section 4.4).

## 5. Summary

A novel beam-profiling system for the RARiS BM4 photon beamline was developed to facilitate quantitative and real-time profiling. The newly developed BPM was incorporated in the system; it employs plastic Sci-Fi and SiPMs as detectors, complemented by a VETO counter designed to remove charged particles and a photon converter enabling profiling of neutral-photon beams. The triggerless DAQ system, streaming TDC, enables the detection of high-intensity photon beams, exceeding several tens of MHz, with position measurements accurate to better than 10 μm for a 1-s beam profile.

Photon-beam measurements were performed under various conditions, revealing that the beam position remains unaffected by changes in beam intensity. However, it was observed that both the beam position and size exhibit a clear correlation with the radiator's position, resulting in time-dependent behavior during a spill.

This newly developed BPM serves as a valuable resource for future physics experiments at the BM4 photon beamline and will contribute significantly to ongoing accelerator research endeavors.

## CRediT authorship contribution statement

**R. Kino:** Writing – review & editing, Writing – original draft, Software, Project administration, Methodology, Investigation, Funding acquisition, Formal analysis, Data curation. **S. Nagao:** Supervision, Software, Resources, Project administration, Methodology, Investigation, Funding acquisition, Formal analysis, Data curation, Conceptualization, Writing – review & editing. **T. Akiyama:** Investigation, Data curation. **H. Fujioka:** Investigation. **T. Fujiwara:** Investigation, Data curation. **T. Ishige:** Software, Investigation, Formal analysis, Data curation. **K. Itabashi:** Investigation, Data curation. **S. Kajikawa:** Methodology. **M. Kaneta:** Supervision, Investigation, Data curation. **M. Mizuno:** Investigation, Data curation. **S.N. Nakamura:** Supervision, Project administration, Funding acquisition, Conceptualization, Writing – review & editing, . **K. Nishi:** Investigation, Data curation. **K. Nishida:** Investigation, Data curation. **K. Okuyama:** Software, Investigation, Formal analysis, Data curation. **F. Oura:** Investigation. **K. Tachibana:** Investigation, Data curation . **Y. Toyama:** Investigation, Data curation. **D. Watanabe:** Investigation, Data curation.

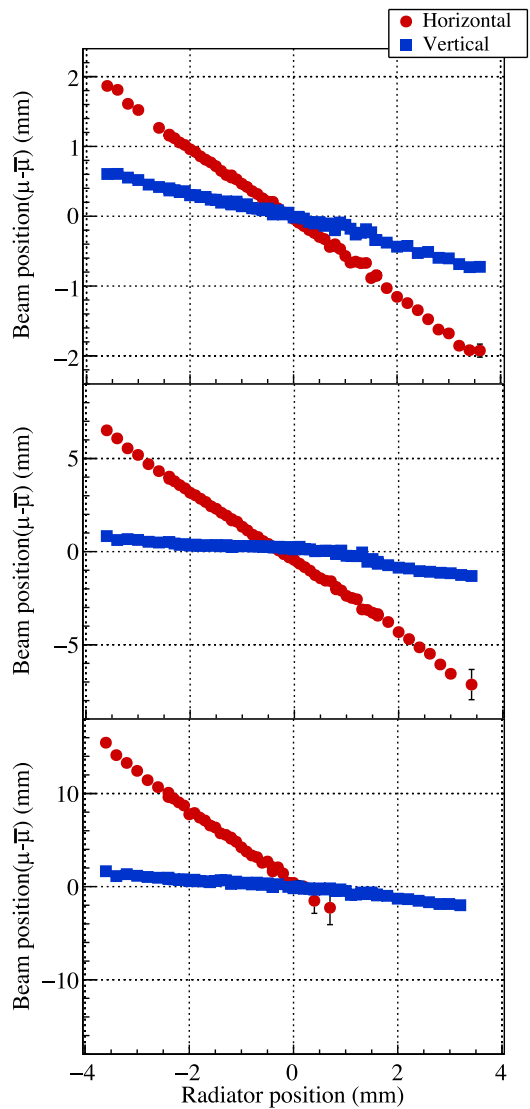

**Fig. 11.** Radiator position dependence of the beam position. The upper figure is measured by BPM1(the direction from the radiator is 2.99 m), the middle is by BPM2(the direction from the radiator is 5.66 m), and the bottom is by BPM3(the direction from the radiator is 10.13 m), respectively. The red points represent the horizontal beam position, and the blue ones represent the vertical beam position. The 0 point on the vertical axis of BPM3 corresponds to the point when the horizontal axis is 0. The range of $x_{rad} > 0.5$ mm is out of the detector's effective area.

**Declaration of competing interest**

The authors declare the following financial interests/personal relationships which may be considered as potential competing interests: Ryoko Kino reports equipment, drugs, or supplies was provided by Research Center for Accelerator and Radioisotope Science at Tohoku University. If there are other authors, they declare that they have no known competing financial interests or personal relationships that could have appeared to influence the work reported in this paper.

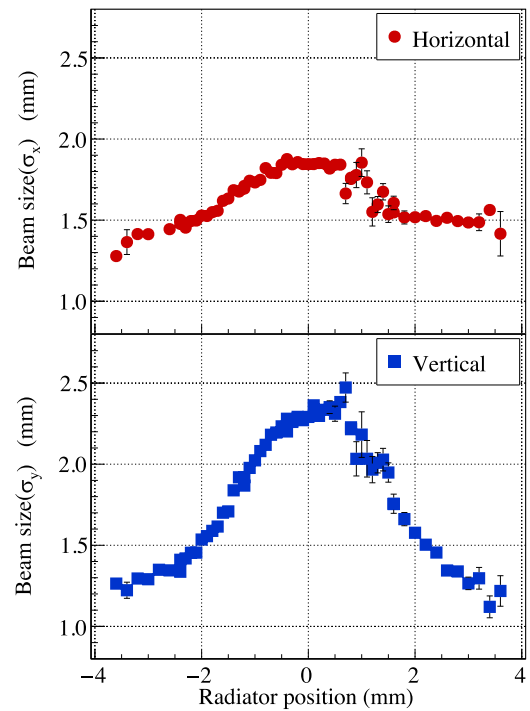

**Fig. 12.** Dependence of the photon-beam size on the position of the radiator at BPM1 (the direction from the radiator is 2.99 m). The red points represent the horizontal beam size, and the blue points represent the vertical beam size.

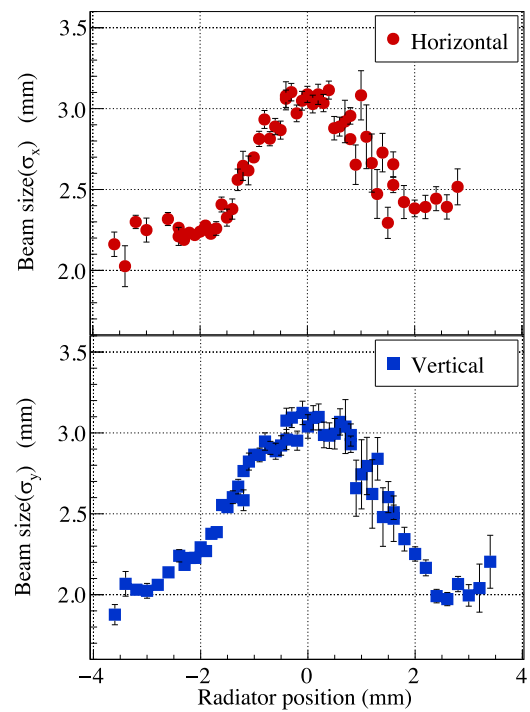

**Fig. 13.** Dependence of the photon-beam size on the position of the radiator at BPM2 (the direction from the radiator is 5.66 m). The red points represent the horizontal beam size, and the blue points represent the vertical beam size.

**Acknowledgments**

The authors appreciate the staff of the RARiS accelerator for providing the primary electron beam and stable operation of the BST ring. In particular, we thank Prof. Fujio Hinode, Prof. Takatsugu Ishikawa, and Dr. Toshiya Muto for their discussions on the accelerator parameters and the behavior of electrons in the BST ring related to this measurement.

This work was conducted under RARiS experiments, Proposal No 2981 and No. 2982.

Financial support for this project was provided in part by JSPS, Japan KAKENHI Grant Number 18H05459, 20H01926, 24H00219, and 24K00657, and Grant-in-Aid for JSPS Fellows, Japan Number 23KJ0180. Five of the authors (RK, SK, KO, FO, and KT) acknowledge support from the JSPS Research Fellowship for Young Scientists, Japan. Seven of us (RK, TA, TI, SK, KO, FO, and KT) is supported by Graduate Program on Physics for the Universe (GP-PU), Tohoku University.

**Data availability**

Data will be made available on request.

## References

[1] F. Hinode, T. Muto, M. Kawai, S. Kashiwagi, Y. Shibasaki, K. Nanbu, I. Nagasawa, K. Takahashi, H. Hama, Upgrade of the 1.2 GeV STB ring for the SR utilization in Tohoku University, J. Phys. Conf. Ser. 425 (7) (2013) 072011.

[2] T.U. Research center for Electron-Photon Science, (URL: https://www.lns.tohoku.ac.jp).

[3] H. Yamazaki, T. Kinoshita, K. Hirota, T. Katsuyama, T. Itoh, A. Katoh, T. Nakabayashi, O. Konno, T. Takahashi, K. Maeda, J. Kasagi, The 1.2 GeV photon tagging system at LNS-Tohoku, Nucl. Instrum. Methods Phys. Res. A 536 (1) (2005) 70–78, http://dx.doi.org/10.1016/j.nima.2004.07.144, URL https://www.sciencedirect.com/science/article/pii/S0168900204016808.

[4] M. Kaneta, B. Beckford, T. Fujii, Y. Fujii, K. Futatsukawa, Y. Han, O. Hashimoto, K. Hirose, T. Ishikawa, H. Kanda, C. Kimura, K. Maeda, S. Nakamura, K. Suzuki, K. Tsukada, F. Yamamoto, H. Yamazaki, Neutral kaon spectrometer 2, Nucl. Instrum. Methods Phys. Res. A 886 (2018) 88–103, http://dx.doi.org/10.1016/j.nima.2017.12.076, URL https://www.sciencedirect.com/science/article/pii/S0168900217314948.

[5] T. Ishikawa, H. Fujimura, H. Fukasawa, R. Hashimoto, T. Ishida, S. Kaida, J. Kasagi, A. Kawano, S. Kuwasaki, K. Maeda, F. Miyahara, K. Mochizuki, T. Nakabayashi, A. Nakamura, K. Nawa, S. Ogushi, Y. Okada, K. Okamura, Y. Onodera, Y. Saito, Y. Sakamoto, M. Sato, H. Shimizu, H. Sugai, K. Suzuki, S. Takahashi, Y. Tsuchikawa, H. Yamazaki, H. Yonemura, The FOREST detector for meson photoproduction experiments at ELPH, Nucl. Instrum. Methods Phys. Res. A 832 (2016) 108–143, http://dx.doi.org/10.1016/j.nima.2016.06.054, URL https://www.sciencedirect.com/science/article/pii/S0168900216305964.

[6] T. Nishizawa, Y. Fujii, H. Kanda, M. Kaneta, Y. Kasai, J. Kusaka, K. Maeda, S. Nagao, S.N. Nakamura, K. Tsukada, F. Yamamoto, Development of a fast timing counter with a monolithic MPPC array, IEEE Trans. Nucl. Sci. 61 (3) (2014) 1278–1283, http://dx.doi.org/10.1109/TNS.2014.2321657.

[7] T. Ishikawa, H. Fujimura, H. Hamano, R. Hashimoto, Y. Honda, T. Ishida, S. Kaida, H. Kanda, S. Kido, Y. Matsumura, M. Miyabe, K. Mizutani, I. Nagasawa, A. Nakamura, K. Nanbu, K. Nawa, S. Ogushi, Y. Shibasaki, H. Shimizu, H. Sugai, K. Suzuki, K. Takahashi, S. Takahashi, Y. Taniguchi, A. Tokiyasu, Y. Tsuchikawa, H. Yamazaki, A fast profile monitor with scintillating fiber hodoscopes for high-intensity photon beams, Nucl. Instrum. Methods Phys. Res. A 811 (2016) 124–132, http://dx.doi.org/10.1016/j.nima.2015.12.027, URL https://www.sciencedirect.com/science/article/pii/S0168900215015983.

[8] Z. Liang, T. Hu, X. Li, Y. Wu, C. Li, Z. Tang, A cosmic ray imaging system based on plastic scintillator detector with SiPM readout, J. Instrum. 15 (07) (2020) C07033, http://dx.doi.org/10.1088/1748-0221/15/07/C07033.

[9] M. Mazziotta, C. Altomare, E. Bissaldi, S.D. Gaetano, G.D. Robertis, P. Dipinto, L. Di Venere, M. Franco, P. Fusco, F. Gargano, F. Giordano, N. Lacalamita, F. Licciulli, F. Loparco, S. Loporchio, L. Lorusso, F. Maiorano, S. Martiradonna, M. Mongelli, F. Pantaleo, G. Panzarini, M. Papagni, C. Pastore, R. Pillera, M. Rizzi, D. Serini, R. Triggiani, A light tracker based on scintillating fibers with SiPM readout, Nucl. Instrum. Methods Phys. Res. A 1039 (2022) 167040, http://dx.doi.org/10.1016/j.nima.2022.167040.

[10] H.U.L. Module, (URL: http://openit.kek.jp/project/HUL). (in Japanese).

[11] R. Honda, T. Aramaki, H. Asano, T. Akaishi, W.C. Chang, Y. Igarashi, T. Ishikawa, S. Kajikawa, Y. Ma, K. Nagai, H. Noumi, H. Sako, K. Shirotori, T. Takahashi, Continuous timing measurement using a data-streaming DAQ system, Prog. Theor. Exp. Phys. 2021 (12) (2021) http://dx.doi.org/10.1093/ptep/ptab128, 123H01, arXiv:https://academic.oup.com/ptep/article-pdf/2021/12/123H01/42899334/ptab128.pdf.

[12] T. Uchida, Hardware-based TCP processor for gigabit ethernet, IEEE Trans. Nucl. Sci. 55 (3) (2008) 1631–1637.

[13] T. Muto, H. Hama, T. Ishikawa, H. Kanda, S. Kashiwagi, F. Hinode, K. Nanbu, I. Nagasawa, K. Takahashi, C. Tokoku, E. Kobayashi, H. Saito, Y. Shibasaki, Impact of Coulomb scattering in radiator wire on Bremsstrahlung gamma ray in an electron storage ring, in: Proceedings of the 12th Annual Meeting of Particle Accelerator Society of Japan, 2015.

[14] T. Muto, Private communication, 2022.